\documentclass[AMA,STIX1COL,doublespace]{WileyNJD-v2}

\articletype{Main Paper}%

\received{26 April 2016}
\revised{6 June 2016}
\accepted{6 June 2016}

\begin{document}

\title{Bias correction in treatment effect estimates following data-driven biomarker cutoff selection}

\author[1]{Chi Zhang}
\author[1]{Wei Shi}
\author[1]{Spencer Woody}
\author[1]{Qing Liu*}

\authormark{ZHANG \textsc{et al}}

\address[1]{\orgdiv{Center for Design and Analysis}, \orgname{Amgen Inc.}, \orgaddress{\state{Thousand Oaks, California}, \country{USA}}}

\corres{*Qing Liu, \email{qliu02@amgen.com}}

\presentaddress{This is sample for present address text this is sample for present address text}

\abstract[Summary]{Predictive biomarkers are playing an essential role in precision medicine. Identifying an optimal cutoff to select patient subsets with greater benefit from treatment is critical and more challenging for predictive biomarkers measured with a continuous scale. It is a common practice to perform exploratory subset analysis in early-stage studies to select the cutoff. However, data-driven cutoff selection will often cause bias in treatment effect estimates and lead to over-optimistic expectations in the future phase III trial. In this study, we first conducted extensive simulations to investigate factors influencing the bias, including the cutoff selection rule, the number of candidates cutoffs, the magnitude of the predictive effect, and sample sizes. Our insights emphasize the importance of accounting for bias and uncertainties caused by small sample sizes and data-driven selection procedures in Go/No Go decision-making, and population and sample size determination for phase III studies. Secondly, we evaluated the performance of Bootstrap Bias Correction and the Approximate Bayesian Computation (ABC) method for bias correction through extensive simulations. We conclude by providing a recommendation for the application of the two approaches in clinical practice. }



\maketitle


\section{Introduction}

Precision medicine has become increasingly important in medical research and clinical development by tailoring treatments to the individual patient based on their disease characteristics and/or biomarker profiles. It has shown particular promise in cancer treatment \cite{precision_oncology2,precision_oncology1} and can be applied across a broad range of health conditions \cite{precision_initiation}. For clinical development, taking patient heterogeneity into account in trial design can significantly increase the probability of trial success, improve efficiency, and demonstrate the treatment effectiveness for the right population. Therefore, adaptive enrichment designs have been popular and powerful, which enable the selection of promising patient subgroups defined by predictive biomarkers during the trial based on accumulating data\cite{adaptive1,adaptive2,adaptive3,adaptive4}. 

Predictive biomarkers that can predict patients' response to a specific treatment play a critical role in precision medicine and enrichment trials. For example, HER2-positive breast cancer patients benefit from trastuzumab\cite{HER2}. For colorectal cancer patients, EGFR inhibitors like cetuximab and panitumumab only have an effect in patients with wild-type KRAS tumors\cite{KRAS}. In immuno-oncology, PD-L1 expression levels are associated with better responses to immune checkpoint inhibitors\cite{PD-L1,PD-L1lung}. Unlike the dichotomous nature of KRAS mutant versus wild-type, biomarkers measured on a continuous scale, such as PD-L1 expression, require an effective quantitative cutoff to accurately identify the subgroup of patients who will benefit most from the treatment. This is crucial when using predictive biomarkers in clinical development and practice.

Numerous studies have been conducted in this field to either identify subgroups based on pre-selected candidate cutoffs or to determine the optimal cutoff from the entire spectrum of continuous biomarker measurements using statistical models or machine learning \cite{bootstrap2014,gotte2017simulation,gotte2020adjustment,Woo2020DeterminationOC}. The objective is to identify the subgroup defined by an optimal cutoff that demonstrates promising treatment effects based on certain predefined decision rules. Subsequently, the identified biomarker cutoff, along with the estimated treatment effects, will be utilized to support future trial designs or to provide evidence to guide clinical practice.

However, only a limited number of studies have discussed the selection bias arising from the intrinsic attributes of these data-driven cutoff selection methods, particularly when employing Maximum Likelihood Estimation (MLE) \cite{bias}. This bias leads to overestimation of treatment effects within the selected subgroup, as the selection process tends to favor cutoffs that show the most promising results in the given dataset. The overestimated treatment effect can mislead subsequent analyses and decision-making and may result in underpowered confirmatory studies with a high failure rate. Therefore, the first objective of this paper is to systematically evaluate and quantify the selection bias by investigating various factors in data-driven cutoff selection procedures, including selection rules, sample size, the number of candidate cutoffs, and the magnitude of the biomarker's predictive effect.

To adjust for the selection bias, \cite{bootstrap2014} proposed the Bootstrap Bias Correction method and \cite{gotte2017simulation,gotte2020adjustment} applied the Approximated Bayesian Computation (ABC) approach. These studies assessed the performance of these approaches through simulations. However, their simulation studies did not focus on biomarker cutoff selection, nor did they consider various factors that could influence the bias of treatment effect estimates in subgroups defined by the selected cutoff. Without systematic evaluations of these bias adjustment methods in the context of biomarker cutoff selection, practical recommendations on the appropriate method to use under different scenarios to effectively reduce bias remain unavailable. Thus, the second objective of our study is to compare the performance of Bootstrap Bias Correction, ABC, and MLE within the context of cutoff selection and to provide recommendations based on our findings.

The rest of the manuscript is structured as follows: In Section 2, we present a statistical setting for our study and provide a detailed introduction of the methods being investigated, including MLE, Bootstrap Bias Correction, and the ABC approach. Section 3 describes the design of the simulation studies for investigating the selection bias in the data-driven cutoff selection procedure and for evaluating the performance of bias correction methods. Section 4 presents the simulation results. The manuscript concludes a discussion in section 5.

\label{sec:intro}

\section{Methods}
\subsection{General Setting}
We consider the experimental treatment E in comparison to the control group C with a focus on the primary binary endpoint with 1:1 randomization. Let's assume there are total $n$ subjects, $y_{i} \in \{0,1\}$ is the observed binary response, and $x_{i}$ is the value of a continuous biomarker for subject $i$ , $i=1,...,n$. Subjects assigned to the experimental treatment are characterized by $m_i=1$, while those assigned to the control group are denoted by $m_i=0$. In this study, we only consider the continuous biomarker as a predictive biomarker and the underlying model of the binary response is the Bernoulli probability function,
\begin{equation}
\begin{split}
f(y_i|x_i,m_i)=p_i^{y_i}(1-p_i)^{1-y_i}
\end{split}
\end{equation}
\begin{equation}
\begin{split}
p_i=\Pr(y_i=1|x_i,m_i)
\end{split}\end{equation}
\begin{equation}
\begin{split}
logit(p_i)=\beta_0+\beta_1 x_i +\beta_2 m_i +\beta_3 x_i m_i.
\end{split}
\end{equation}

In this study, we use the difference in objective response rates (ORR) between experimental treatment and control to measure the treatment effect; the larger the ORR difference, the better the treatment effect.

We evaluate the primary endpoint within subsets defined by specific cut-off values of a continuous biomarker. Assume there are $K$ candidate cutoff values, $c_1$,..., $c_K$,  the subset $s_k$ is defined as the group of subjects with biomarker larger than $c_k$, where $s_k=\{i| x_i> c_k\}$ and $k=1,2,...,3$, thus there are in total K subsets. Among the $K$ subsets under consideration, one subset is chosen based on a predetermined selection rule that considers the difference in ORR between the experimental treatment and control groups across subsets. Let $\theta_k=p_{Ek}-p_{Ck}$  ($k=1, \ldots, K$) be the true treatment effect in subset $k$, where $p_{kE}$ and $p_{kC}$ are the ORRs in subset k for experimental and control group, and let $\theta_{select}$ be the true treatment effect in the selected subset. The procedure of how to obtain the true treatment effect under our simulation studies with continuous biomarkers is recorded in Section~1 of our Supplementary Document.

To evaluate the selection bias under different biomarker cutoff selection scenarios and procedures, we apply the commonly used MLE to estimate the treatment effect. The estimator of treatment effect in the selected subset is denoted by $\hat{\theta}_{selectMLE}$. Then we evaluate two bias adjustment approaches by using MLE as a benchmark. We consider Approximate Bayesian Computation (ABC) and Bootstrap Bias Correction and use $\hat{\theta}_{selectABC}$ and  $\hat{\theta}_{selectBootstrap}$ to represent the adjusted estimates within the selected subset from the ABC and Bootstrap Bias Correction, respectively. 

In our study, the selection bias is measured by conditional bias, which is defined as the mean of the estimated ORR difference minus the true ORR difference, given the selection of a particular subset. Assuming a total of N simulations, the conditional bias, given the selection of subset k, is expressed as:

$$\Delta_{MLE|s_k \ selected}=\frac{1}{N_k}\sum_{j:s_k \ selected}(\hat{\theta}_{selectMLE,j}-\theta_{select,j}),$$
$$\Delta_{ABC|s_k \ selected}=\frac{1}{N_k}\sum_{j:s_k \ selected}(\hat{\theta}_{selectABC,j}-\theta_{select,j}),$$
$$\Delta_{Bootstrap|s_k \ selected}=\frac{1}{N_k}\sum_{j:s_k \ selected}(\hat{\theta}_{selectBootstrap,j}-\theta_{select,j}),$$where $N_k$ is the number of simulations where $s_k$ is selected and $j$ is the index of $j$th simulation. Table S1 provides a summary of the notations and their descriptions used throughout the paper. Further details of the methods of MLE, Bootstrap Bias Correction, and ABC will be presented in Sections 2.2, 2.3, and 2.4, respectively.

\subsection{Maximum Likelihood estimator}

For each subset k, we calculate the MLEs of the ORR difference between experimental treatment and control group as $\hat{\theta}_{kMLE}= \frac{n_{kER}}{n_{kER}+n_{kEN}}-\frac{n_{kCR}}{n_{kCR}+n_{kCN}}$, where $(n_{kER},n_{kEN},n_{kCR},n_{kCN})$ are the number of subjects in the experimental group with/without response and the number of subjects in the control group with/without response. Then according to a predefined select rule, one of the K MLEs is selected as $\hat{\theta}_{selectMLE}$.

\subsection{Bootstrap Bias Correction}
In this section, we explain how to use the Bootstrap bias correction technique to adjust the bias of the effect estimator of the selected subgroup\cite{bootstrap2014}. For subjects in both experimental treatment and control groups, we collect the response $y_{i}$ and biomarker values $x_{i}$, where  $i=1,...,n$ . The responses $y_{i}$ are assumed to be independently distributed between experimental treatment and control groups, each with its cumulative distribution function $F_E(x_{i}|m_i=1)$ and $F_C(x_{i}|m_i=0)$. In order to evaluate the treatment effect on individuals exhibiting biomarker values greater than the cutoff $c_k$, we define the parameter $\theta_k=t(F_E, F_C,c_k)$ as the true treatment effect among individuals with biomarker greater than $c_k$. An estimator for this parameter is denoted by $\hat{\theta}_k = h(\mathbf{y},\mathbf{x},c_k)$, where $\mathbf{y}$ and $\mathbf{x}$ represent the vectors for response and vectors for biomarker values. The bias of the estimator of $\theta_k$ is then calculated using the formula below.

\begin{equation}
\begin{split}
B_{F_E,F_C}(\hat{\theta}_k,\theta_k) = E_{F_E,F_C}[h(\mathbf{y},\mathbf{x},c_k)]-t(F_E, F_C,c_k)
\end{split}
\end{equation}
where $E_{F_E,F_C}[\cdot]$ represents the expectation to the distributions $F_E$ and $F_C$.

Then with K candidate cutoffs, we define $\theta_{select}$ as the treatment effect of the selected subset based on a predefined selection rule and denote $\hat{\theta}_{select}=s(\mathbf{y},\mathbf{x},c_k,k=1,...,K)$ an estimator of the treatment effect $\theta_{select}$ of the selected subset. We introduce functions $u_k:\mathbb{R}^{n} \times \mathbb{R}^{n}\rightarrow \{0,1\}$  where $\sum_{k=1}^K u_k\leq 1$, to represent the predefined selection rule. These functions determine that at most only one subset can be selected,  with the sunset $k$ being chosen if $u_k(\mathbf{y},\mathbf{x},c_k,k=1,..,K) = 1$.  The conditional bias of $\hat{\theta}_{select}$ given subset $k$ has been selected can be defined as:
    \begin{equation}
    \begin{split}
        B_{F_E,F_C}(\hat{\theta}_{select}|u_k=1) &= 
        \frac{ E_{F_E,F_C}[s(\mathbf{Z})] -t(F_E,F_C,c_k)E_{F_E,F_C}[u_k(\mathbf{Z})]}{E_{F_E,F_C}[u_k(\mathbf{Z})]}\\
        &=
        \frac{ E_{F_E,F_C}[s(\mathbf{Z})] }{E_{F_E,F_C}[u_k(\mathbf{Z})]}-t(F_E,F_C,c_k)\\
        &=
        \frac{ E_{F_E,F_C}[\hat{\theta}_{select}] }{E_{F_E,F_C}[u_k(\mathbf{Z})]}-\theta_{k},
    \end{split}
    \end{equation}
where $\mathbf{Z}=(\mathbf{y},\mathbf{x},c_k,k=1,..,K)$.

Then $F_E$ and $F_C$ should be substituted by $\hat{F}_E$ and $\hat{F}_C$ to obtain the bootstrap estimator of conditional bias. However, the bootstrap estimator of bias \( B_{\hat{F}_E,\hat{F}_C} \) cannot be derived in a closed form for general distributions \( F_E \) and $F_C$, so it is necessary to approximate it via Monte Carlo simulations. To this end, we generate \( B \) Bootstrap samples \( \mathbf{M_{b}^*} = (m_{1b}^*, \ldots, m_{nb}^*) \) for treatment/control group indicator and \( \mathbf{X_{b}^*} = (x_{1b}^*, \ldots, x_{nb}^*) \) for biomarker values and obtain  \( \mathbf{Y_{b}^*} = (y_{1b}^*, \ldots, y_{nb}^*) \) for response values from the estimated distributions \( \hat{F}_E \) and  \( \hat{F}_C \), where $b=1,...,B$. The approximation of the bootstrap estimator of conditional bias is given by:

\begin{equation}
B_{\hat{F}_E,\hat{F}_C}(\hat{\theta}_{select}|u_k=1) \approx \hat{h}_k^* - t(\hat{F}_E,\hat{F}_C,c_k),
\end{equation}
where
\begin{equation}
\hat{h}^*_k = \frac{\sum_{b=1}^B s(\mathbf{Z^*_b})u_k(\mathbf{Z^*_b})}{\sum_{b=1}^B u_k(\mathbf{Z^*_b})}
\end{equation}
and $\mathbf{Z^*_b}=(\mathbf{Y_{b}^*},\mathbf{X_{b}^*},c_k,k=1,...,K)$.

Consequently, we set the Bootstrap Bias correction estimator on the treatment effect of the selected subset as $\hat{\theta}_{selectBootstrap}$  and with a plug-in estimator when subset k is selected, $t(\hat{F}_E,\hat{F}_C,c_k)=\hat{\theta}_{select}$,  $\hat{\theta}_{selectBootstrap}$ given subset k is selected can be expressed as:
\begin{equation}
    \begin{split}
        \hat{\theta}_{selectBootstrap} | (u_k=1) &= \hat{\theta}_{select}-B_{\hat{F}_E,\hat{F}_C}(\hat{\theta}_{select}|u_k=1)\\
        &\approx \hat{\theta}_{select}-\hat{h}_k^* + t(\hat{F}_E,\hat{F}_C,c_k)\\
         &\approx 2\hat{\theta}_{select}-\hat{h}_k^* 
    \end{split}
\end{equation}
where $\hat{\theta}_{select}$ is obtained by MLE in our simulation studies, thus in our simulation $\hat{\theta}_{select}=\hat{\theta}_{selectMLE}$

\subsection{Approximate Bayesian Computation (ABC)}

In data-driven cutoff selections to identify an unknown optimal cutoff, the likelihood function can be analytically intractable. In such cases,  ABC becomes a valuable method in Bayesian statistics to address this issue \cite{Beaumont2002ApproximateBC}. Among the various implementations of the ABC algorithm, the foundational one is the ABC rejection algorithm, which is applied in our study\cite{Sunnker2013ApproximateBC,Beaumont2019ApproximateBC}.

When applying the ABC rejection algorithm, one initial and most critical step is how to select appropriate prior distributions. The priors should ideally cover the true treatment effect with a high probability. When existing knowledge about the parameters of interest is available, this information can be incorporated into the prior. In our study, we start by using the true parameters as the mean of normal distributions to represent this knowledge suggested by previous corresponding studies\cite{gotte2017simulation,gotte2020adjustment},
$$\beta_{00}\sim N( \beta_0, 1/5), \ \beta_{01}\sim N(\beta_1, 1/5),$$
$$\beta_{02}\sim N( \beta_2, 1/5), \ \beta_{03}\sim N( \beta_3, 1/5),$$
where $\beta_{00}$, $\beta_{01}$, $\beta_{02}$, and $\beta_{03}$ are the priors of each parameter.

However, in the absence of strong prior information, non-informative priors are often employed. Common choices include uniform distributions and normal distributions with a wide range. For this study, we select a standard normal distribution as our non-informative prior,
$$\beta_{00}\sim N(0,1), \ \beta_{01}\sim N(0,1),$$
$$\beta_{02}\sim N(0,1), \ \beta_{03}\sim N(0,1).$$
We also consider an alternative approach to derive the priors directly from the observed data. Specifically, we conduct the logistic regression analysis on the observed data, using the coefficients and their standard errors to shape the prior distribution,
$$\beta_{00}\sim N(\hat{\beta}_0,se(\hat{\beta}_0)), \ \beta_{01}\sim N(\hat{\beta}_1,se(\hat{\beta}_1)),$$
$$\beta_{02}\sim N(\hat{\beta}_2,se(\hat{\beta}_2)), \ \beta_{03}\sim N(\hat{\beta}_3,se(\hat{\beta}_3)),$$
where $\hat{\beta}_0$, $\hat{\beta}_1$, $\hat{\beta}_2$ and $\hat{\beta}_3$ are the coefficients of parameters $\beta_0$, $\beta_1$, $\beta_2$ and $\beta_3$ from the logistic regression.

We evaluate the performance of ABC using these three distinct sets of priors. After selecting the priors, $N_{simABC}$ sets of parameters are drawn from the priors, where $N_{simABC}$ is the number of ABC simulations. For each set of parameters, a specific model will be simulated to generate the response, biomarker, and treatment data. If the observed and simulated data are identical, the simulation provides an unbiased estimate of the likelihood and is accepted. However, in practical applications, it is highly unlikely that the observed and simulated data will be identical. To address this issue, researchers typically use summary statistics to represent the observed and simulated data. Ideally, these summary statistics should be sufficient; however, in most cases, such sufficient statistics do not exist. Therefore, the summary statistics should at least be informative. Motivated by previous studies \cite{gotte2017simulation,gotte2020adjustment}, we apply stricter summary statistics, using the response rates for the full population and each subset in both experimental treatment and control groups as our summary statistics. 

Then the summary statistics from the observed data and the simulated data are compared. If the summary statistics are sufficiently close to each other according to the predefined rejection or acceptance rule, the corresponding ABC simulation is accepted; otherwise, it is rejected. The posterior distribution of parameters is approximated from the distribution of parameter values from the accepted ABC simulations.

Motivated by the previous study\cite{gotte2017simulation}, we describe the ABC algorithm used in this paper with K candidate cutoffs $c_k$, where $k=1,...,K$, and assume $c_{select}$ is the cutoff selected in the observed dataset. 
\begin{enumerate}
    \item Simulate $(\beta_{00}^j, \beta_{01}^j,\beta_{02}^j,\beta_{03}^j)$ from one of three sets of priors described above with $j=1,..,N_{simABC}$.
    \item Simulate the n-dimensional vector of biomarker $\mathbf{x}^j$ from Beta(1,1) and the n-dimensional vector of treatment indicator $\mathbf{m}^j$ from Bernoulli(0.5). After that simulate the n-dimensional vector of the response $\mathbf{y}^j$ from equation (1) $f(.|\mathbf{x}^j,\mathbf{m}^j)$ with $j=1,..,N_{simABC}$.
    \item Calculate the summary statistics $s^j=S(\mathbf{x}^j,\mathbf{y}^j,\mathbf{m}^j,c_k,k=1,..,K)=(s_{full,E}^j,s_{full,C}^j,s_{subset \ k,E}^j, s_{subset \ k,C}^j, \ k =1,...,K )$, where $s_{full,E}^j$ and $s_{full,C}^j$  represent the response rates for the full population in the experimental treatment and control groups , and $s_{subset \ k,E}^j$ and $s_{subset \ k,C}^j$ represent the response rates for the subset $k$ in the experimental treatment and control groups for $j$-th ABC simulation.
    \item If $S_{obs,full,E}\in[s_{full,E}^j-\epsilon,s_{full,E}^j+\epsilon]$, $S_{obs,full,C}\in[s_{full,C}^j-\epsilon,s_{full,C}^j+\epsilon]$and $S_{obs,subset \ k,E}\in[s_{subset \ k,E}^j-\epsilon,s_{subset \ k,E}^j+\epsilon]$ and $S_{obs,subset \ k,C}\in[s_{subset \ k,C}^j-\epsilon,s_{subset \ k,C}^j+\epsilon]$for every $k=1,...,K$, then $w^j=1$ otherwise $w^j=0$, where $S_{obs, full,E}$, $S_{obs, full,C}$ , $S_{obs, subset \ k,E}$ and $S_{obs, subset \ k,C}$ are the summary statistics from the observed dataset, representing the response rates for the full population and subset k in the experimental treatment and control groups.
    \item Calculate the true treatment effect in the subset with cutoff $c_{select}$ and define it as $\theta_{select}^j$ with $j=1,...,N_{simABC}$.
\end{enumerate}
Then the ABC-adjust estimate $ \hat{\theta}_{selectABC}$ in the selected subset is obtained by the median of the posterior distribution $\pi_{ABC}(\theta_{select}|S_{obs})$ which is $\inf_m\{\frac{1}{\sum_{v}^{N_{simABC}}w^v} \sum_{j=1}^{N_{simABC}}w^jI_{\theta_{select}^j\leq m}\geq 0.5\}$.

\section{Simulation Studies}

\subsection{Bias in data-driven biomarker cutoff selection}
In this section, we conducted extensive simulations to evaluate the bias in data-driven biomarker cutoff selection with various scenarios and selection procedures. In each simulation setting, we generated random datasets with $n$ subjects with $\frac{n}{2}$ subjects in the experimental treatment group and $\frac{n}{2}$ subjects in the control group (1:1 randomization). The treatment variable $m_i$ was sampled from Bernoulli(0.5), so if they are from the experimental treatment group then we set $m_i=1$ otherwise we set $m_i=0$. We considered one continuous biomarker, and generated the quantile  $x_{i}$ for each subject from Beta(1,1) to reflect the actual practice of using quantile in the analysis as the original value of a continuous biomarker is often highly skewed distributed. The biomarker is independent of treatment assignment. Then the response for each individual was generated from a Bernoulli distribution with the probability as:

$$\frac{exp(\beta_0+\beta_1x_{i}+\beta_2m_i+\beta_3x_{i} m_i)}{1+exp(\beta_0+\beta_1x_{i}+\beta_2m_i+\beta_3x_{i} m_i)}$$
Since this study focuses on predictive rather than prognostic biomarkers, for simplicity, $\beta_1$ was set to zero. 

Different factors were considered in the simulation studies to investigate different data-driven cutoff selection designs, including 1) sample size, 2) number of candidate cutoffs, 3) magnitude of the predictive effect of the biomarker, and 4) selection rule as summarized below:
\begin{itemize}
    \item Sample sizes: 20, 40, 60, and 100 subjects per arm.
    \item Numbers of candidate cutoffs: two cutoffs (0.3 and 0.6); and five cutoffs (0.2, 0.3, 0.4, 0.5, and 0.6). Each cutoff responds to the subset with subjects whose biomarker values are larger than that cutoff. For example, two subsets (subjects with a biomarker larger than 0.3 and subjects with a biomarker larger than 0.6) will be used as two candidate subsets in the simulation study with two cutoffs. 
    \item Magnitudes of the predictive effect: different $\beta_3$ settings.
    \begin{itemize}
        \item more or less (1): $\beta_0 = - 0.4$; $\beta_1 = 0$; $\beta_2 = 0.2$; $\beta_3 = 0.2$.

        \item more or less (2): $\beta_0 = - 0.4$; $\beta_1 = 0$; $\beta_2 = 0.2$; $\beta_3 = 0.5$.
    \end{itemize}
    \item Selection rules:  Selection Rule 1 is based on the maximum observed treatment effect and Selection Rule 2 is based on posterior probabilities. The details of these two selection rules will be explained below.
\end{itemize}

We chose two different values of  $\beta_3$ to explore the influence of the magnitude of the predictive effect on bias correction. Based on equation (3), the interaction term, $\beta_3$, captures how the treatment effect depends on the biomarker value. Thus, a higher $\beta_3$ implies a stronger interaction between the biomarker and the treatment effect, resulting in a more pronounced linear increase in the treatment effect as the biomarker value increases. From Table S2, we can also observe the average treatment effects, calculated as the difference in response rates between the experimental and control groups, are larger in the simulation scenario with larger $\beta_3$ (more or less (2)) compared to the scenario with a smaller $\beta_3$ value (more or less (1)), consistently across the full population and all subsets.

Under Selection Rule 1, we first calculate the ORR difference in each subset and select the subset with the largest estimated ORR difference between experimental and control groups. In Selection Rule 2, we begin by calculating the posterior probability that the ORR difference is larger than 0.15 for each subset. Subsets with a posterior probability greater than 0.7 are considered candidate subsets. If more than one subset meets the criterion, the subset with the largest prevalence is selected. For example, if both subsets corresponding to cutoffs of 0.3 and 0.6 satisfy the condition, we select the subset with cutoff 0.3 since there are more subjects included in the subsets with biomarker values larger than 0.3 compared to 0.6. If no subset meets the posterior probability criteria, no subset will be selected.  

\subsection{Performance of selection bias adjustment methods}

In the second part of the simulation studies, we aim to evaluate the performance of Bootstrap Bias Correction and ABC in adjusting for selection bias by using MLE as the benchmark. The comparison between Bootstrap Bias Correction and MLE was conducted under a broader set of simulation settings, considering two different magnitudes of the predictive effect and four varying sample sizes, while maintaining the same two candidate cutoffs (0.3 and 0.6). Given the computational demands for the ABC algorithm, the simulations for this method were restricted to more or less (1) scenario with two candidate cutoffs (0.3 and 0.6) and a sample size of 40 subjects per arm. For both Bootstrap Bias Correction and ABC methods, we evaluated the performance under two different selection rules mentioned above.

\section{Results}

\subsection{Bias in data-driven biomarker cutoff selection}
We first evaluated the bias of the estimated treatment effect by MLE under the first selection rule with two candidate cutoffs by selecting the cutoff leading to the largest estimated treatment effect. As shown in Figure 1, MLE tends to overestimate the treatment effects when a cutoff of 0.6 is selected. This overestimation is more pronounced with a smaller predictive effect (more or less 1) or smaller sample size. In contrast, when a lower cutoff of 0.3 is selected, MLE maintains its unbiased nature. Upon further analysis, when cutoff 0.3 is selected we observe a tendency to overestimate treatment effects for subjects with the biomarker value between 0.3 and 0.6 and to underestimate treatment effects for subjects with the biomarker value larger than 0.6. This equilibrium in biases across different subsets contributes to the unbiased nature of MLE for treatment effect when a 0.3 cutoff is selected (Fig S1).

\begin{figure}
    \centering
    \includegraphics[width=0.75\linewidth]{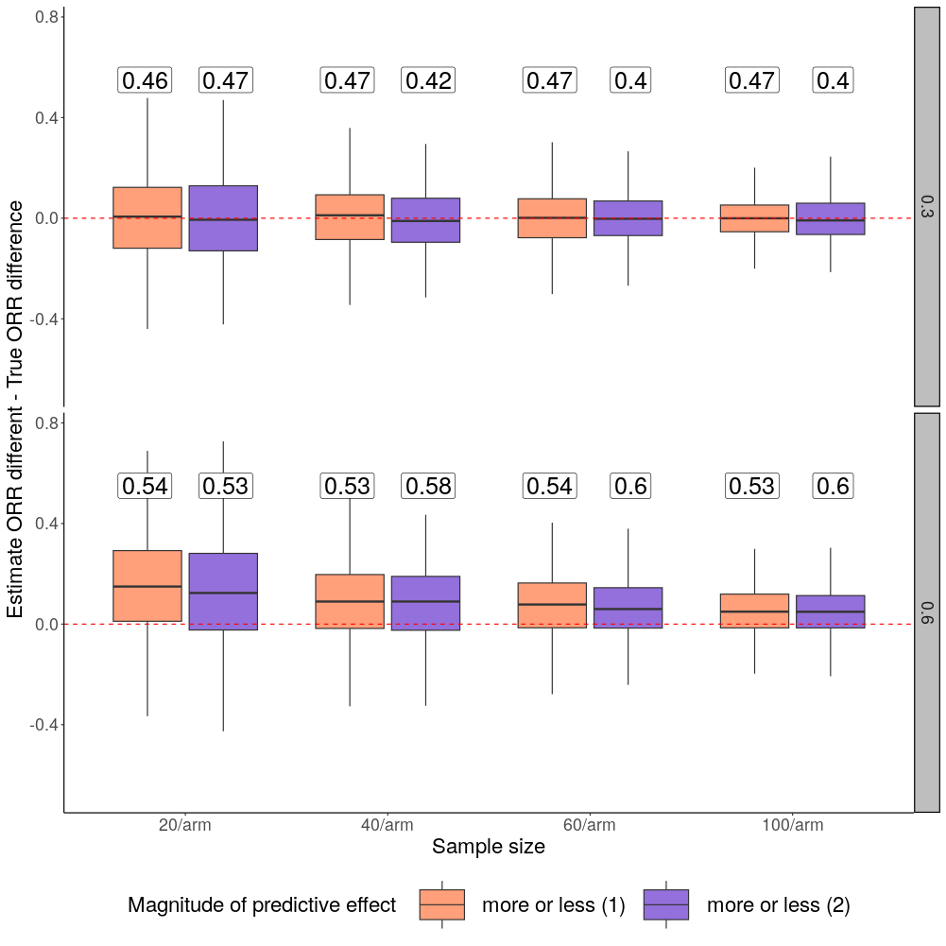}
    \caption{\textbf{Conditional bias under the first selection rule with two candidate cutoffs. }The x-axis is the sample size and the y-axis shows the conditional bias (Estimated ORR difference - True ORR difference given selected cutoff). The first row represents the conditional bias when a cutoff of 0.3 is selected, and the second row is for 0.6. The numbers above each boxplot are the probabilities of each cutoff selected. The dashed line represents zero. Different color boxes represent different magnitudes of effect on the predictive biomarker.}
    \label{fig:enter-label}
\end{figure}

Subsequently, we expanded the range of candidate cutoffs from two to five, specifically including 0.2, 0.3, 0.4, 0.5, and 0.6. We observe a consistent pattern that a smaller predictive effect is associated with a larger bias (Figure S2). Additionally, it becomes evident that the larger the value of selected cutoffs, the larger the biases. This phenomenon can be attributed to the natural characteristics of the first selection rule and the positive effect of the predictive biomarker, where larger cutoffs correspond to smaller selected sample sizes, but with subjects exhibiting larger treatment effects. Notably, when the lowest cutoff, 0.2, is selected in this scenario, MLE is also almost unbiased. Besides, when we compare the two scenarios with different numbers of candidate cutoffs, we also observe that the more candidate cutoffs for selection, the larger the estimation bias when the same cutoff is selected (Figure S3). 

Then we assessed the bias of using MLE in biomarker cutoff selection under the second selection rule based on posterior probability and prevalence considered. Specifically, if multiple subsets exhibit a posterior probability of ORR difference larger than $15\%$ greater than 0.7, the subset with the highest prevalence will be selected.  By subtracting the response rate in the control group from that in the experimental treatment group (Table S2), we found that the treatment effect exceeds 0.15 only in the subset with a biomarker value greater than 0.6 under the more or less (2) scenario. Thus, most simulation trials result in no subsets being selected under this scenario with the second selection rule (Figure S4). Furthermore, since the second selection rule aims to identify subsets that meet the minimum sufficient benefit with higher prevalence, the smaller cutoff is more frequently selected than the larger cutoff (Figure 2).

\begin{figure}
    \centering
    \includegraphics[width=0.75\linewidth]{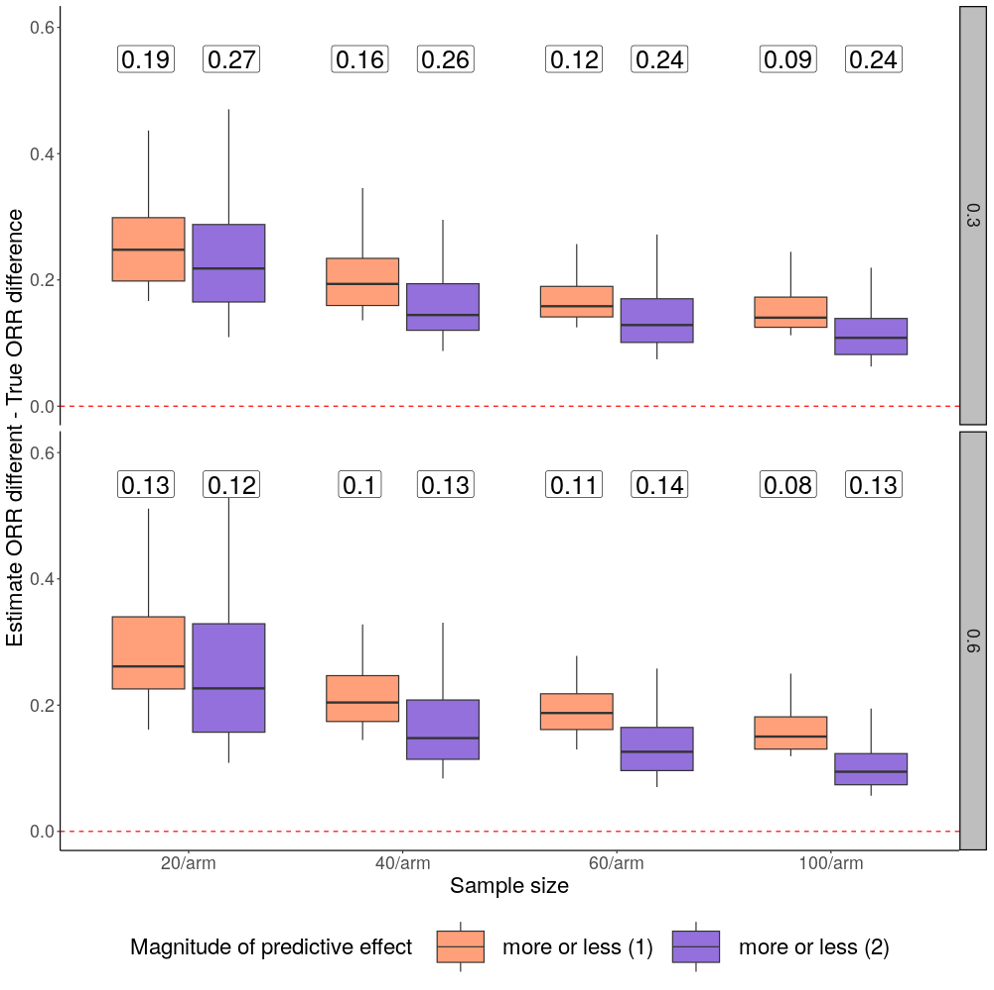}
    \caption{\textbf{Conditional bias under the second selection rule with two candidate cutoffs. }The x-axis represents the sample size and the y-axis shows the conditional bias (Estimated ORR difference - True ORR difference given selected cutoff). The first row represents the conditional bias when a cutoff of 0.3 is selected, and the second row is for 0.6. The numbers above each boxplot are the probabilities of each cutoff selected. The dashed line represents zero. Different color boxes represent different magnitudes of effect on predictive biomarkers.}
    \label{fig:enter-label}
\end{figure}

With two candidate cutoffs, we observed that MLE consistently overestimates the treatment effect, regardless of the selected cutoff, which is different from the results when using the selection rule 1 (Figure 2). However, consistent with the results under the first selection rule, a decrease in predictive effect or a smaller sample size results in increased estimation bias. Besides, the second selection rule introduces a more pronounced bias compared to that in the first selection rule and exhibits selection bias no matter which cutoff is selected. When comparing the bias of the same subset selected across different numbers of candidate cutoffs, we observe an intriguing trend. Unlike our finding under the first selection rule, the second rule shows that the more cutoffs available for selection, the lesser the estimation bias when focusing on the same subset (Figure S5). This can be attributed to the inherent characteristics of the second selection rule. When the minimum sufficient benefit criterion is satisfied, the cutoff corresponding to the larger prevalence is chosen, which results in a smaller estimation bias for the 0.3 cutoffs when selecting from five cutoffs compared to two. A similar rationale can also be applied to the 0.6 cutoff. Thus, the impact of the number of cutoffs on estimation bias varied by the selection rule and the specific cutoff under consideration.

In conclusion, the magnitude of selection bias is contingent on the cutoff selection rule. Even though in certain instances, MLE can be unbiased, the bias exists in most scenarios and the magnitude is not negligible. Moreover, the larger the sample size and the predictive effect of the biomarker, the smaller the selection bias in the treatment effect estimate within the selected subset. However, the influence of the number of candidate cutoffs is not uniform, depending on different cutoff selection rules.

\subsection{Performance of selection bias adjustment methods}

The above simulation results indicate that estimates derived from MLE may be prone to an upward bias due to the data-driven selection procedure. This hidden bias can create an overly optimistic view, making certain subsets appear more promising than they might be in reality. The estimation bias could lead to an incorrect decision to select these subsets as the target population for subsequent confirmatory trials. Furthermore, overestimating the treatment effect can impact the sample size calculation, potentially resulting in underpowered Phase III studies. Therefore, accounting for the selection bias and uncertainties arising from smaller sample sizes is crucial. We performed simulation studies to compare the performance of Bootstrap Bias correction and the ABC approach to the selection bias adjustment. 

\subsubsection{Comparison between Bootstrap Bias Correction and MLE}

We compared the performance of the Bootstrap Bias Correction with MLE with two candidate cutoffs. Given the first selection rule, when cutoff 0.6 is selected, we observe that Bootstrap Bias Correction can significantly reduce the bias from MLE estimates and this bias decreases as the the sample sizes increase (Figure 3B). When a cutoff of 0.3 is selected, Bootstrap Bias Correction provides an unbiased estimation, aligning with the unbiased nature of the MLE in this scenario (Figure 3A). Under the second selection rule, similarly, the Bootstrap Bias Correction reduces the bias for both cutoffs and the bias decreases as the sample size and the predictive effect increases (Figure S6 and S7).

\begin{figure}
    \centering
    \includegraphics[width=0.75\linewidth]{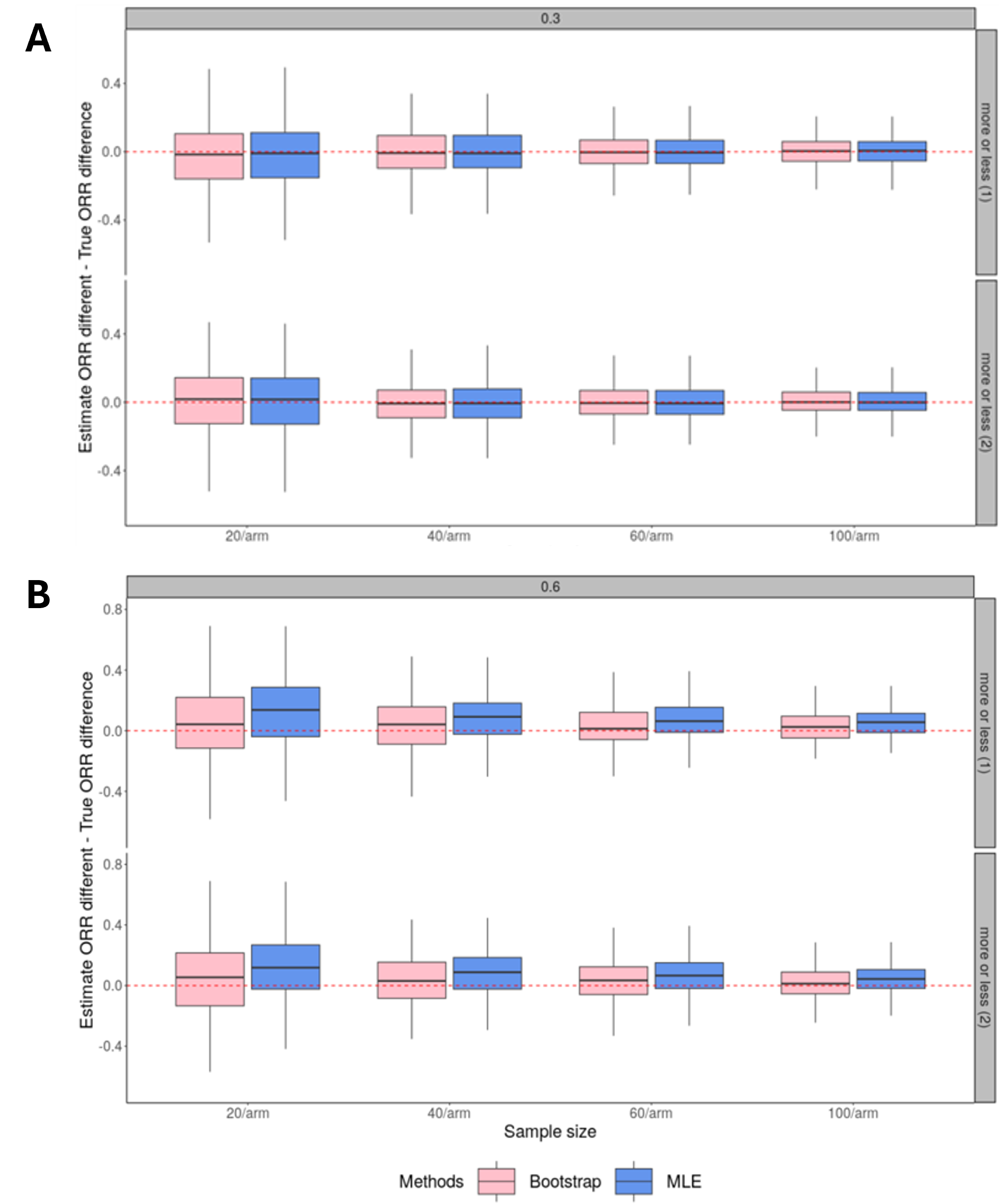}
    \caption{\textbf{Conditional }\textbf{bias for both Bootstrap Bias Correction and MLE under the first selection rule with two candidate cutoffs when cutoff (A) 0.3 is selected or (B) 0.6 is selected.} The x-axis represents the sample size and the y-axis shows the conditional bias (Estimated ORR difference - True ORR difference given selected cutoff). The first row represents the conditional bias when more or less (1) model is used, and the second row is for more or less (2) model. The dashed line represents zero. Different color boxes represent different methods.}
    \label{fig:enter-label}
\end{figure}

\subsubsection{Comparison between ABC and MLE}

In Section 2.4, we introduce three distinct setups of prior distributions in ABC. The first prior utilizes the true parameter values as the means of normal distributions, with a variance set to $0.2$. The second prior uses standard normal distributions to represent a non-informative prior. Lastly, the third prior is constructed using coefficients and their associated standard errors estimated from logistic regression. These three priors are referred to as "ABC\_true", "ABC\_standard", and "ABC\_logit", respectively.

We compare the performance of ABC under different priors with MLE (Figure S8). The choice of prior has a substantial impact on the performance of the ABC. Notably, the prior based on the true parameter values shows superior performance. However, in practical applications, obtaining such a prior will not be feasible. The use of a non-informative prior demonstrates a certain capability in adjusting bias. Nevertheless, it is essential to acknowledge the proximity of the true beta parameters to zero in our model, particularly for $\beta_1$, which is set to zero. This proximity may result in favorable performance for "ABC\_standard" due to coincidental alignment with the true parameter values. Therefore, a sensitivity analysis is conducted to ensure robustness against variations in prior assumptions.

In the sensitivity analysis, we shift the true $\beta_1$ further from zero to 0.3. The results reveal that when using non-informative prior (standard normal distribution), ABC does not consistently yield reliable estimates (Figure S9).  The sensitivity of the results to variations in prior assumptions highlights the importance of selecting appropriate priors in ABC, especially in the absence of strong prior information. This analysis indicates that in practical applications, where prior information is uncertain, careful consideration should be given to alternative priors to ensure robustness. Thus, in the subsequent section comparing ABC with Bootstrap Bias Correction, the priors constructed using the coefficients from logistic regression are used for the ABC algorithm.

\subsubsection{Comparison among MLE, Bootstrap Bias Correction, and ABC}

We conducted a comparative analysis of MLE, Bootstrap Bias Correction, and ABC algorithm under the more or less (1) scenario with 40 subjects per arm. Given the first selection rule, both Bootstrap Bias Correction and ABC can reduce the bias; however, Bootstrap Bias Correction demonstrates superior bias correction than ABC (Figure 4A). Under the second selection rule, Bootstrap Bias Correction consistently reduces the bias across all selected cutoffs (Figure 4B). In contrast, ABC only reduces the bias when the cutoff 0.6 is chosen, making ABC less effective (Figure 4B). Overall, Bootstrap Bias Correction provides a more accurate bias-adjusted estimation, even though it leads to larger standard errors of bias compared to both ABC and MLE (Figure 4A, 4B).

\begin{figure}
    \centering
    \includegraphics[width=1\linewidth]{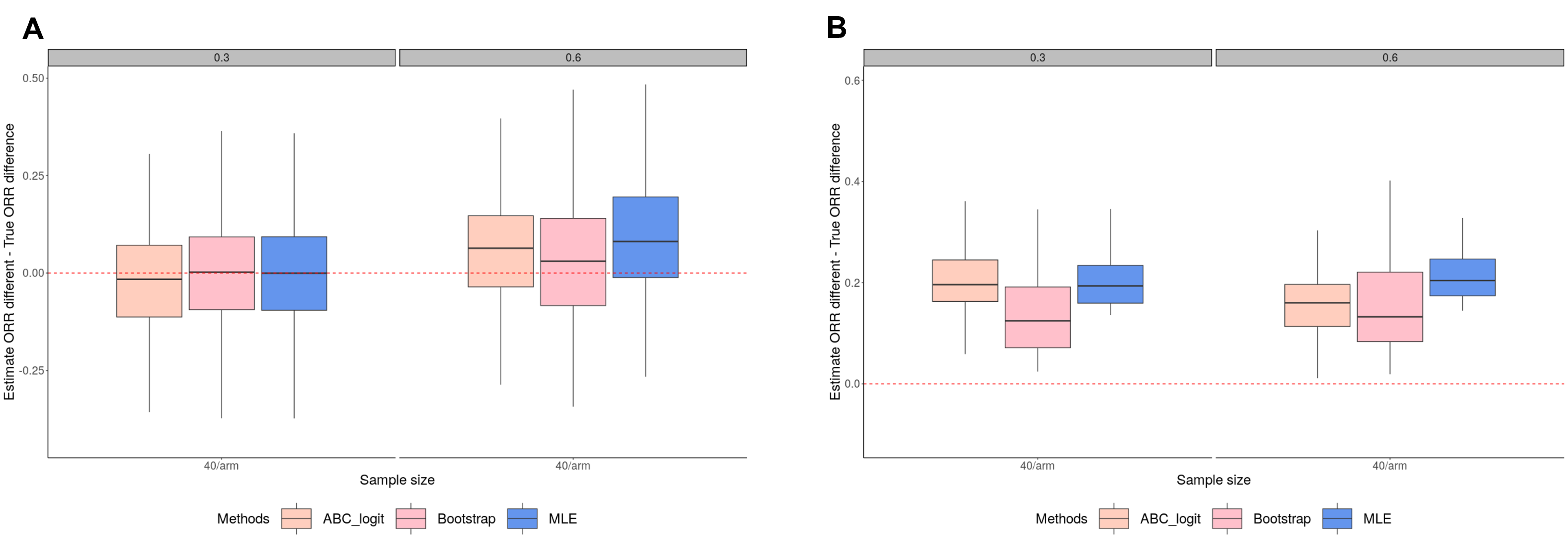}
    \caption{\textbf{Conditional bias for ABC, Bootstrap Bias Correction, and MLE with two candidate cutoffs under (A) the first selection rule and (B) the second selection rule. }The x-axis represents the sample size and the y-axis shows the conditional bias (Estimated ORR difference - True ORR difference given selected cutoff). The first column represents the conditional bias when cutoff 0.3 is selected, and the second column is for cutoff 0.6. The dashed line represents zero. Different color boxes represent different methods.}
    \label{fig:enter-label}
\end{figure}

\section{Discussion}
As far as we are aware, this is the first study to systematically assess the selection bias associated with using MLE for biomarker cutoff selection utilizing data-driven selection criteria across various scenarios. Our simulation studies elaborated that the magnitude of the selection bias can vary substantially depending on the selection rule and the number of candidate cutoffs. Notably, as the sample size and the magnitude of the predictive effect decrease, the bias tends to increase. For example, in one of our simulations, when the true treatment effect was set at 0.09 and the decision boundary was assumed at a 15\% response difference, applying MLE overestimated the treatment effect to 0.18 in the selected subset. This overestimation could lead to misguided decision-making, pushing for further analysis under the expectation of over-optimized results. Our findings emphasize the importance of considering bias when using data-driven cutoff selection procedures based on MLE. An overestimated treatment effect can potentially misinform clinical practice, lead to incorrect Go/NoGo decisions, select suboptimal patient populations, or result in underpowered confirmatory studies, which could hinder the efficiency of drug development for precision medicine.

To mitigate the risk, we evaluated the performance of selection bias adjustment approaches in the context of biomarker cutoff selection using the data-driven procedure. Bootstrap Bias correction consistently reduced selection bias across all simulation scenarios. The correction of the treatment effect estimate can decrease the probability of making a wrong Go decision and optimize the phase III study design with a higher probability of success. While a small amount of bias has remained, the magnitude is negligible. Compared to Bootstrap Bias correction, the ABC approach can also adjust for the selection bias, however, its performance is highly sensitive to the choice of prior distributions. Under certain circumstances, it did not perform as expected. Considering the infeasibility of specifying correct prior distribution in practice and the high computation burden required by ABC, Bootstrap Bias correction is recommended.

Looking forward, there are several avenues for future work. One key area is the development of methods to adjust the confidence intervals in the presence of bias correction. Additionally, different ABC algorithms, such as ABC MCMC and ABC SMC can also be evaluated in the future. Overall, our study provides comprehensive simulations to investigate the bias in treatment effect estimates following data-driven biomarker cutoff selection and compares different bias correction methods, emphasizing the importance of adjusting for selection bias in biomarker-driven clinical trials and providing practical recommendations.

\bibliography{WileyNJD-AMA}%




\end{document}